# The AMI Database Project: Atlas Data Challenge Bookkeeping, and the Tag Collector, a new tool for Release Management.

Solveig Albrand, Johann Collot, Jerôme Fulachier.
*LABORATOIRE DE PHYSIQUE SUBATOMIQUE ET DE COSMOLOGIE,
IN2P3-CNRS/ Université Joseph Fourier,
53, avenue des Martyrs, 38026 Grenoble Cedex,
France.*

Many database tools have the same or similar requirements. The Atlas Metadata Interface (AMI) project aims to provide a set of generic tools for managing database applications. AMI has a three-tier architecture with a core that supports a connection to any RDBMS using JDBC and SQL. The middle layer assumes that the databases have an AMI compliant self-describing structure. It provides a generic web interface and a generic command line interface. The top layer contains application specific features. Currently 7 such applications exist. Two of these applications are described. The first, and principal use of AMI, is the Atlas Data Challenge Production Bookkeeping interface. The second application is called Tag Collector, a web tool for release management, has many features which have greatly facilitated Atlas software management.

## 1. INTRODUCTION

The "Atlas Metadata Interfaces" (AMI) project [1] started in the spring of 2000 with the requirement to provide an electronic notebook for the Atlas Liquid Argon sub detector test beam acquisition. The application catalogues information about run data using a relational database. Three user interfaces were provided for this application; a GUI, a web interface and a C++ API as part of the Atlas Athena Framework.

Other database applications were rapidly requested from the same developers for projects with very similar requirements. In particular, the interface requirements are often almost the same; all projects require an efficient web interface for searching; many projects require a command line interface or a C++ or Java API. Evidently, it makes sense to reuse as much work as possible, and this implies that the architecture of the software must allow development to be generic.

This paper describes the architecture which we have chosen and also the two main applications of the software used within the ATLAS experiment; the AMI Production Database currently used for Atlas data challenges, and the Tag Collector, which is a collaborative tool for release management.

## 2. THE ARCHITECTURE OF AMI

### 2.1. Principles

The principles that have guided our choice of architecture are:

- A relational database is used.
- The software should be independent of the particular RDBMS used,
- It should be possible to manage database schema evolution.
- The system should support geographic distribution,
- The interfaces should be as generic as possible,
- The software should not depend on a particular operating system.

Our choice of a relational database was dictated both from the fact that development time is minimized when learning time is minimized, and also by the lack of conviction that any currently available technology will provide better searching efficiency.

It seemed to us very important to remain independent of a particular relational database, for several reasons. Firstly, although we chose mySQL initially because of the extreme rapidity with which a web interface can be built, using PHP, this database lacks, or lacked at the time we began the project, many features which are fairly standard in its peers, such as record level transactions. Secondly we need to consider the scalability of a free database such as mySQL or PostGres, as compared to the well-known power of Oracle. Our desire to provide a tool which could be geographically distributed is a third motivation for multi RDBMS support. Large computing centres that own licenses and have knowledge of, and manpower to support, a database such as Oracle, are by and large extremely unwilling to diversify into providing support for another database. On the other hand, smaller less rich centres will not be able to use a tool which depends on buying an expensive license. So, our architecture does not only permit different RDBMS to be used, it permits them to be used within the application at the same time.

Geographic distribution is a desirable feature in any tool provided for the Atlas collaboration, which is, of course, itself widely distributed. Our tools should be available at all times in a robust, reliable way. We need to bear in mind that if a tool is to be adopted successfully within Atlas, all features must be able to be scaled up. We have decided to work towards the "data warehousing" model [2] – with several source databases, allowing concurrent input, and central read-only databases updated automatically, which permit complex requests, without affecting the writing efficiency. Such a model has other advantages, for example, an individual production site could have complete access rights on a part of the database, and declare that it is ready for uploading to the main database, only when the data has been validated. It should also be pointed out that in this model, different schema could be employed in the source database and the central database. A source database schema should be optimized for the simplicity of input, whereas the database to which queries will be addressed will be optimized for the efficiency of these queries.

As mentioned above, generality and abstraction are pre-requisites for reuse. We decided to base our software on a sub set of SQL, and to make our databases self-describing. More explicitly, each AMI compliant database contains its own description, in terms of the entities it contains, and their relationships. All our interfaces are designed to exploit these descriptions.

Schema evolution is a complex database problem to which it is impossible to do justice in this short article. Suffice it to say that all methods of managing this problem are based on database modelling techniques [3]

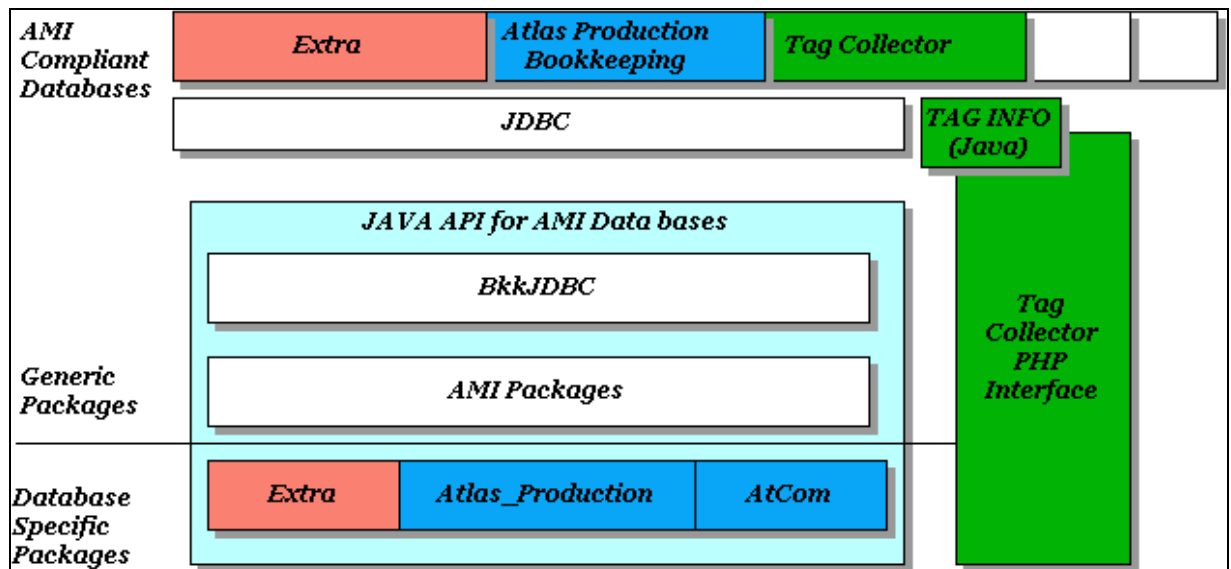

Figure 1: A Schematic View of the Software Architecture of AMI

The user does not connect directly to a database, but to a "router" database. The router will direct the user connection to the correct database which is defined in terms of the user parameters The user does not need to know the name of the database consulted, or on which server it is situated, or which RDBMS is used by that server.

The software is implemented in JAVA, which of course allows requirement 5 to be satisfied.

## 2.2. Software Architecture

Figure 1 shows the architecture of AMI. The core layer is an abstract layer based on JDBC which manages all the database connections, and wraps the basic SQL requests.

The middle layer manages the AMI compliant databases in a generic way, using the in-built database descriptions. In this layer, no assumptions are made about the names of databases, their contained entities or the relations between the entities.

Only three databases are shown in the figure, but currently 6 projects use the AMI base classes. Four of them are in use for Atlas; the original LAr bookkeeping interface, Shaper which is used for electronics quality control, and the two applications which are described in more detail below. One other is used for simulation Bookkeeping for the Gedeon Reactor Physics project.

The outer software layer has application specific software. The application specific software in most cases consists of a specialized web page, built on the generic functions. This is the case for Atlas Production. The AtCom [4] packages, built on AMI, implement a graphic interface for the control of production jobs. The "External Traffic Analyser" (EXTRA) project is also worthy of note. This application is used in the context of the IN2P3 network security monitoring. It uses threads to scan router activity at fixed intervals. Other threads correlate the data from the different routers, and store it in "fact" tables optimized for analysis, thus implementing the "warehouse" model described above. The results are displayed graphically, and systems managers can be warned by e-mail if unusual or suspect activity is observed.

## 2.3. Deployment

Figure 2 shows the deployment of the AMI databases on different servers. The client software connects firstly to a router database. In function of the configuration of the client, the connection is referred to the correct bookkeeping database. Bookkeeping databases may be distributed geographically, and may be running with different RDBMS. Router databases will be replicated automatically.

## 2.4. Logical Organization of the Data

As mentioned above, data in AMI databases is organized in projects. Each project can contain several processing steps. The data that is described in one project can undergo several processing steps, in a predefined order. A processing step maps to a physical database. Several projects of the same type can share the same processing steps. Each processing step contains a certain number of "entities". An entity maps to a database table. The list of entities, and the relations between them are described within the processing step database. Other database information informs the AMI generic interfaces of the behaviour of a particular field, for example, whether the parameter is part of the default set used for the AMI default query, or gives the user information about the meaning of a field, or the units in which the value is expressed.

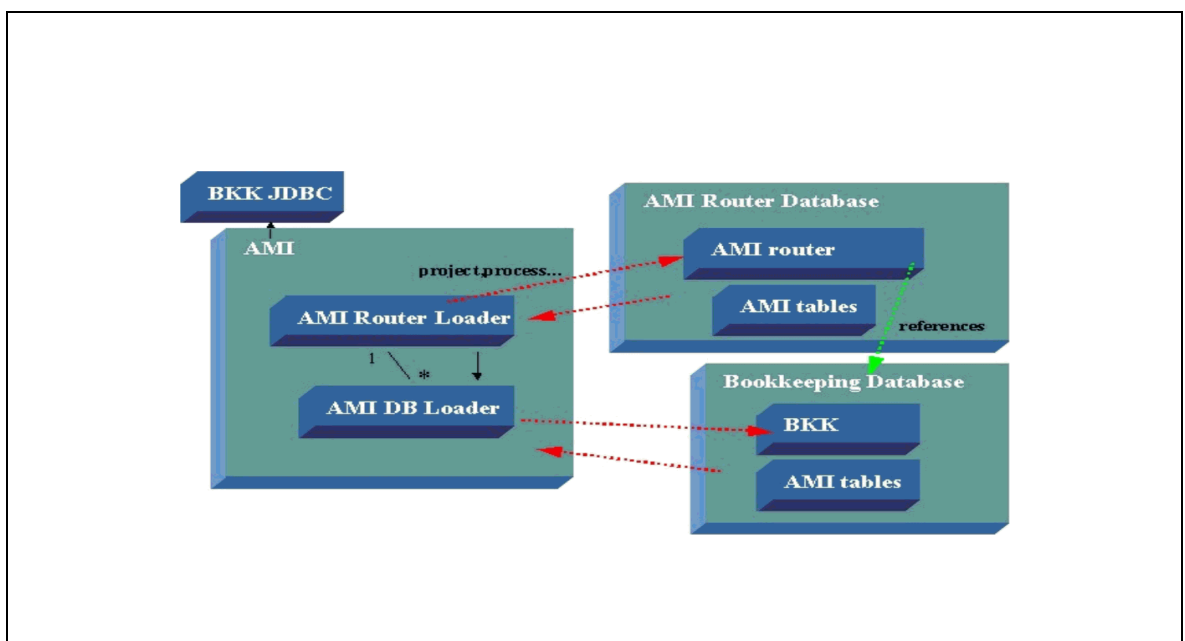

Figure 2: The Deployment of AMI on Several Servers

## 3. EXAMPLE 1 – ATLAS PRODUCTION BOOKKEEPING

This project has the potential to be the major user of AMI in the future, both by the size of the databases involved, the number of potential users, and by the certainty that the database requirements will evolve during the lifetime of the application. Therefore, although it is not at present the most used of our databases, it is a driving force in the design.

The aim of the Atlas Production Bookkeeping is to provide a link from physics criteria used in a production job to the logical file names of the files which contain the data. In the terms of Grid Architecture, AMI is an "application meta data base" [5]. AMI does not manage any data concerning the physical location of data or any data replication. Atlas Data Challenge 1 uses the MAGDA tool [6] for this function. The AMI web interface links to MAGDA to obtain physical location data.

The database currently occupies 31 MB of disk space; we estimate that it will grow to 1.2 GB over the next three years.

### 3.1. Database Schema

The following is a glossary of some of the terms used in defining the schema for the AMI Atlas production database. It should be noted that these are not official Atlas definitions!

**event** : The ensemble of data for a particular beam crossing, or a subset of this data. Event data may be "real", directly recorded from the detector for a particular set of trigger conditions, or simulated, using Monte Carlo techniques.

**dataset :** A collection of events.

**dataset number: :**A integer tag, which is assigned to a dataset. This number is analogous to the run number assigned by the DAQ in the case of real data.

**partition :** A file which contains a part of a dataset. Datasets have to be divided into partitions because of file size limitations.

**partition number :** An integer, from 1 to N, where N is the number of partitions created for a given dataset.

**project :** A set of datasets which have been created with the same physics, or computing purpose. Each project has a project name, for example "dc0", "dc1".

**processing step :** A dataset, once created, may undergo a sequence of different processes. We refer to each process in the sequence as a processing step. Each processing step has a name, for example "simul". Different projects may choose to define different sets of processing steps. A processing step maps to a particular algorithm or sequence of algorithms. Processing steps are applied to datasets in a predefined order; e.g. Simulation always comes BEFORE Reconstruction and AFTER Event Generation.

**logical dataset name :** A dataset name is a combination of other terms, which is unique within Atlas. For example "projectName.datasetNumber.processingStep".

**logical file name :** A tag which completely identifies a partition. It must be unique within the Atlas collaboration. It consists of at least, the dataset Name and the partition number.

**dataset derivation :** Since within a given project, results of one processing step will be used as input data for the following processing steps, it is possible to trace the ascendants and descendents of a particular dataset within the project.

**Figure 3** shows a partial schema of the AMI database. Process, project and project type are generic types which are defined by the user's configuration. The router will direct the users connection to the corresponding bookkeeping database. Each database contains a certain number of entities. These elements, and their relations are described within the table, so the software to manage them can be generic. The figure shows the entities for Atlas production processes. Also shown are two specific Atlas Production elements, determined by the project type. These are related to all the datasets of all Atlas production processes.

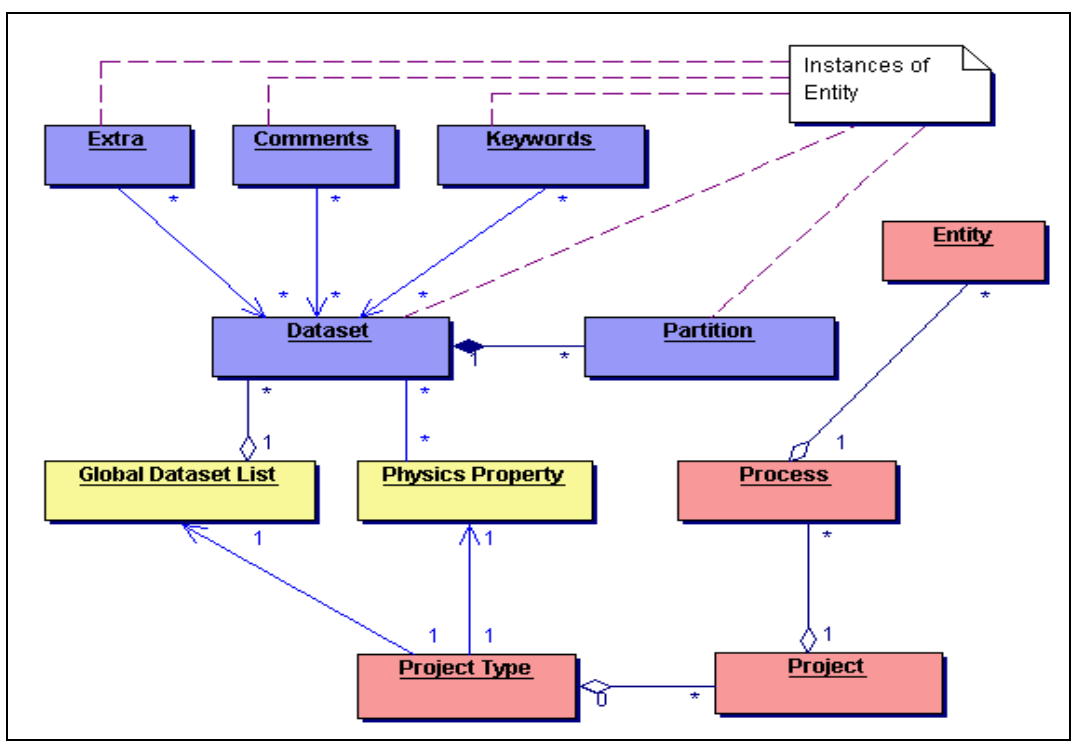

Figure 3 : Overview of the AMI Atlas Production Database Schema

## 3.2. Web Interface

### 3.2.1. Generic

Figure 4 shows a screen shot of the AMI Quick Search web interface. This interface is (almost) completely generic, and will work for any AMI compliant database.

### 3.2.2. Application Specific

Figure 5 shows the result of a query for a particular dataset. A text box shows the SQL query which was used, and it can be edited and re-submitted if desired. By clicking on the logical dataset name, a graph which shows both the "parent" and "children" datasets of the selected dataset. The result page also permits navigation to the "partitions", which make up the dataset (not visible in the figure).

### 3.3. Current Status

AMI has been largely used for Atlas Data Challenge 1. Currently 60000 files are catalogued for three processing steps, representing about 500 datasets. Data has been entered from about 40 different sites.

There are about ten users of the web interface a day.

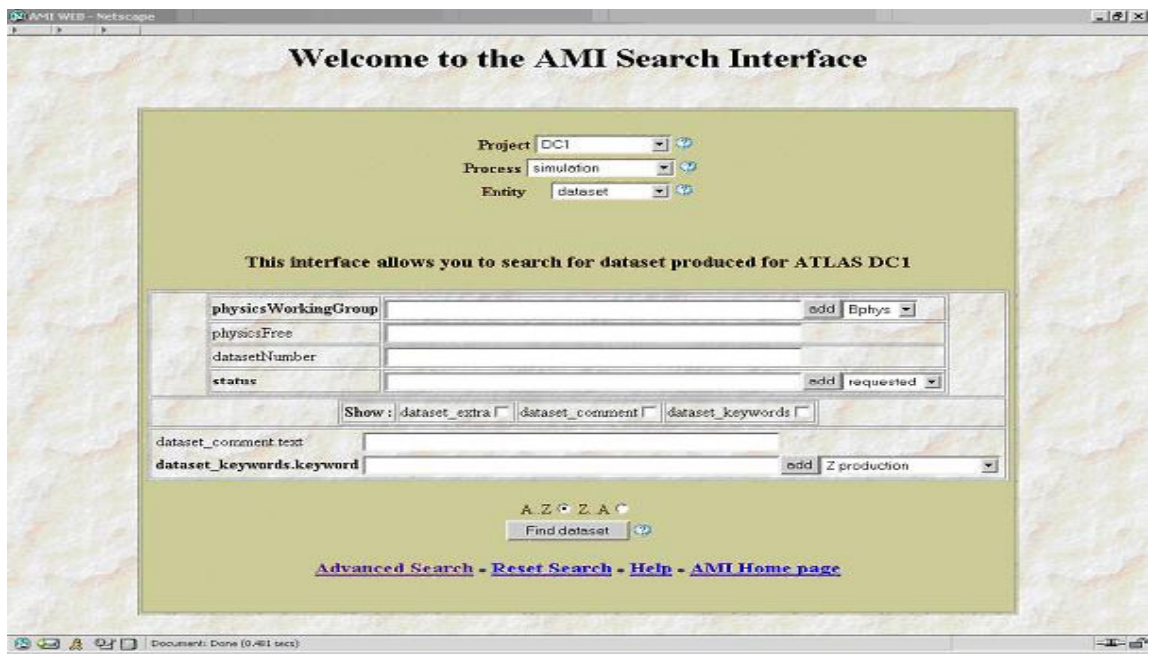

Figure 4 : A screenshot of the AMI Quick Search Interface

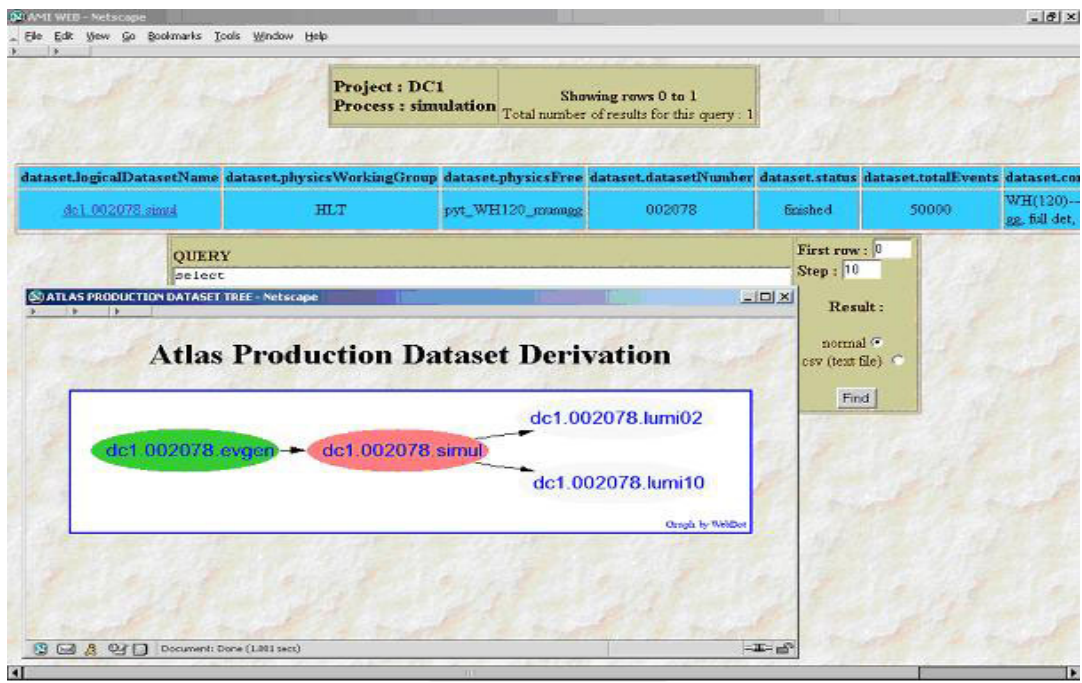

Figure 5 The result of a query for dataset dc1.002078.simul, and the graph of the derivation. The graph allows navigation between different processing steps.

### 3.4. Future Plans

In the near future we plan to continue to enrich the AMI web pages with new features. An early priority is to add some access control so that write functions can be implemented. Three classes of functions will be provided

Functions for Atlas Physicists to allow the adjunction of text comments to a dataset, or to add an extra attribute value pair to a dataset. These two functions are already available using the command line interface.

Functions for project managers, to allow some limited modification of schema, and also to set up some special queries, for example, giving an overview of the state of advancement of production.

Functions for AMI database administrators to facilitate the configuration of an AMI compliant database.

Longer term plans include the implementation of web services to facilitate communication with other applications, in particular grid middleware. We plan to integrate a Spitfire interface to AMI [7]

### 4. EXAMPLE 2 – TAG COLLECTOR

Tag collector is a tool designed for assisting in the management of Atlas software releases. The database schema is relatively simple compared to the above example. However, the application has become very important to Atlas developers and release managers alike and the site [8] averages around 50 hits a day. Developers have found it greatly simplifies the inclusion of their packages in a release, and release managers and librarians appreciate that it enforces a certain discipline among developers. As we are unaware of another tool with the same functionality, we will describe the tool's functions in some detail.

### 4.1. The Atlas Release Problem

Atlas offline software – like that many other projects, includes a large number of packages, developed by a large number of people, distributed geographically over several time zones.

The software components are interdependent, and in some cases very strongly coupled. These packages must be built together.

Table 1 gives an idea of some of the numbers involved.

Table 1: Size of the Atlas Release

| | |
|---|---|
| Number of software packages | 570 (of these 300 contain C++ source code) |
| Number of developers | 107 |
| Number of C++ source files | ~2000 |
| Number of SLOC | ~$10^6$ |

The strategies which must be put in place to manage this code are mostly dependent on developer goodwill. All developers must place their code in the same CVS repository, using the same directory structure. Developers follow a fairly straight forward policy for tagging successive version of their packages. Packages are organized hierarchically using the notion of "containers".

Atlas uses the configuration management tool CMT [9] since the end of 2001. The CMT environment supports naming conventions for packages and directories, and provides tools for automating as much as possible the implementation of these conventions. It permits the recursive description of configuration requirements and it can automatically deduce from the description the effective set of configuration parameters needed to build packages or to use them.

The Atlas release is described by a cascade of "requirements files", each one describing which other packages are needed to use the package in question. The top requirements file is a list of all the container packages that make up the release. One person – the Atlas librarian- is responsible for establishing this top list, and to do this he or she must rely in information given by the various container packages concerning the correct package tag to be included. Package container managers are in the same way, reliant on information from the actual code package developers. Source code package developers on the other hand need to know when a different package which they use has changed version. This situation was classically managed in Atlas by a cascade of e-mails, echoing the package dependency chain, and of course increasingly complicated as the number of packages grows, subject to all sorts of human errors, and inefficient in our multi-time zone environment. In addition, we had no structure for management of the contents of a release. It was all too easy for a prolific developer to introduce a well-meaning change in his package just before a build, often with unsuspected border effects. Developers were also asking for regular, and frequent developer builds, so in April 2002,the first trials of a system for regular nightly builds were put in place [10].

During the summer of 2001, the situation was reaching a crisis point. The complexity of the release procedure, and the fact that no user support was available for the release tool in use at the time forced developers, and management to introduce a more rigid control. It was decided that one physicist would take the responsibility for coordinating the objectives of each successive release. This was the trigger for the first version of Tag Collector, which was released to developers at the end of August 2001.

Figure 6 shows the activity of the Atlas software developers' list, correlated with the introduction of various tools.

## 4.2. Tag Collector; a Database Solution

The introduction a database to hold the ensemble of developer tags required for a release has been a great step forward in controlling the release procedure. Developers now declare their new tags using the web interface to Tag Collector. The librarian and the release coordinator have special privileges in controlling the actions of the developers.

Although keeping a centralized record of the tags which were included in successive releases is in itself useful for orderly development, the main advantage of Tag Collector is that it allows the enforcing of a number of rules.

Here are some of the main features of the tool:

1) When new release is opened, all the tags of the previous release are taken to initialize the tag tree. If nothing is done, the new release gets build with the tags of the previous release.

2) The developers working on sub-packages use Tag Collector to declare the new tags they would like to see included in the release in progress. They can no longer declare a non-existent tag, because the tool collects the existing tags from the CVS repository, and presents them in a combo-box.

3) The container package managers are warned by e-mail that developers have added new tags to their contained packages. They know that they must alter the container requirements files, and re-tag the container packages. This is done before a deadline fixed by the release coordinator.

4) Once the container manager has tagged the container, no further sub-package tags can be declared unless the concerned container package managers agree to open the sub-package tag collection again for the release in progress.

5) The tool verifies the CMT requirements files for container packages, in function of the declared sub-package tags, and will not permit tagging if an error is found.

6)   The librarian is able to obtain from the tool a complete list of correct tags to make the new release.

The release coordinator is allowed to impose locks on packages to prevent unwanted new tags being included in the release. In particular, a certain number of packages are declared to be "core" packages, because they are used by most other packages. They are unlocked and locked independently from the other packages, and this feature imposes a stability of core packages during the pre-release period, so that other developers have the possibility to test their own new software versions.

Tag Collector provides a web service notably used by the scripts that prepare the automatic nightly builds. Individual developers can also use the service to query the database for example to obtain the list of packages in a release, or to obtain the name of the package manager.

Figure 7 gives an overview of the architecture of the Tag Collector tool.

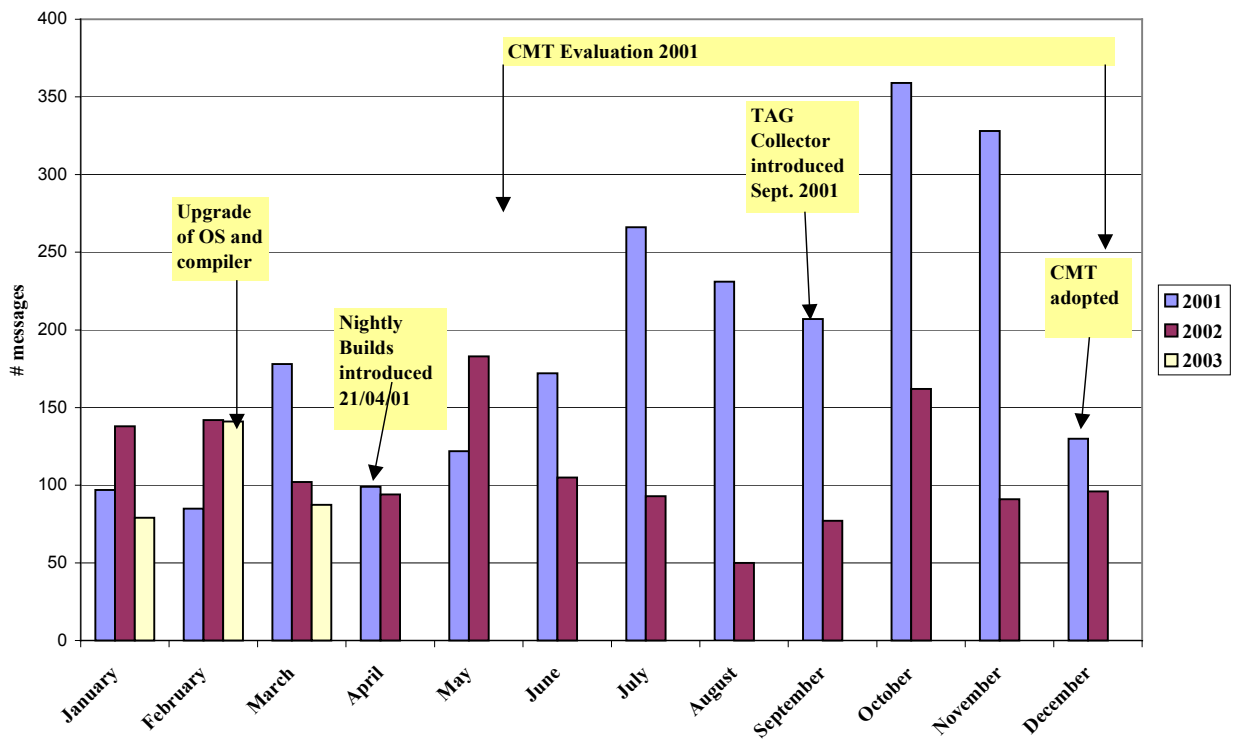

Figure 6 : Activity on the Atlas Software Developers' Mailing List.

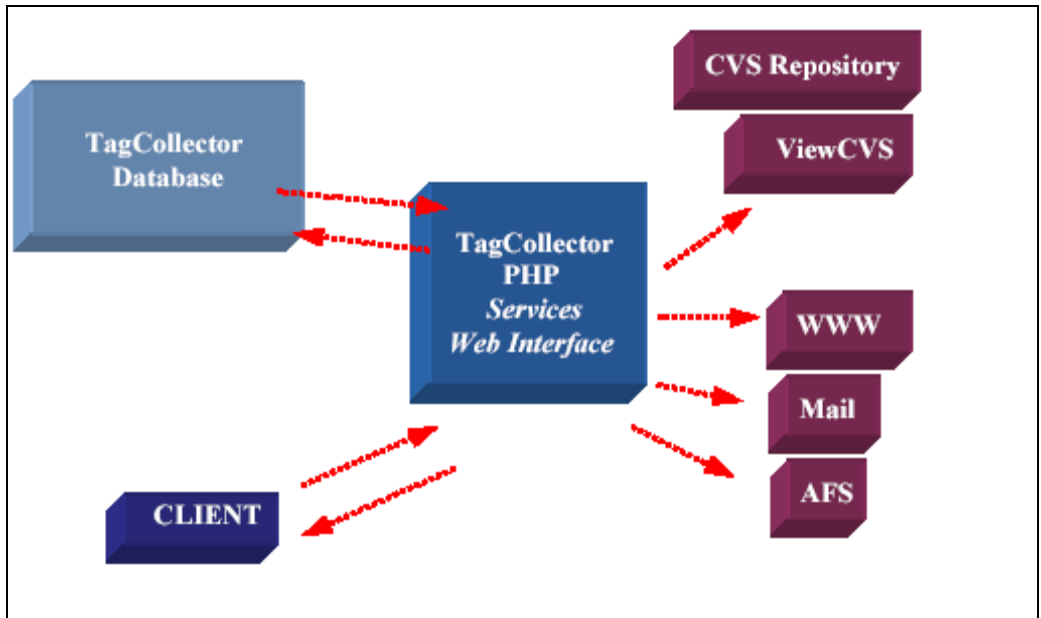

Figure 7 : Schematic View of Tag Collector.

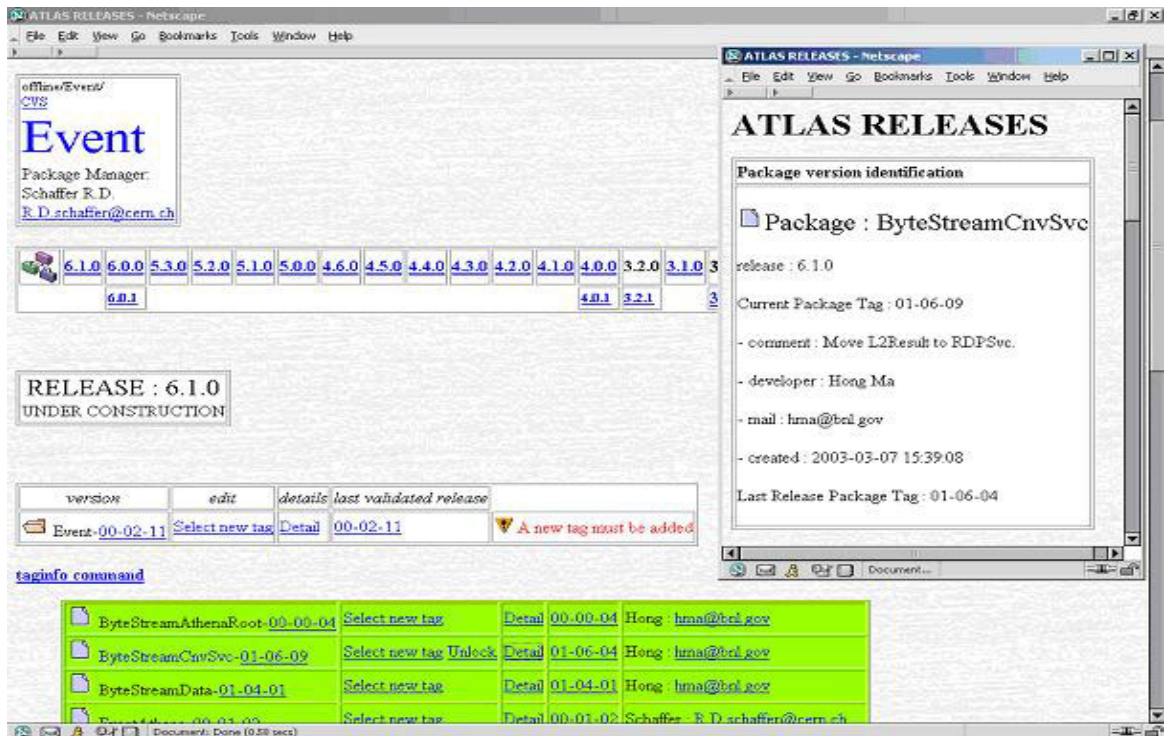

Figure 8 : A screenshot of the Tag Collector.

Figure 8 is a screenshot of the Tag Collector web interface for one particular package. The release tree allows the user to navigate in the different reviews for a particular package. This screen shot shows the package manager that the container package must be retagged because one of the developers has modified some code. The modified package, *ByteStreamCnvSvc*, is shown as locked, and one can see both the new tag (01-06-09) and the last valid tag (01-06-04). Superimposed is a part of the "detail" for this package, where we can see from the comment what the developer did in his modification. Lower down on the detail page – and invisible in the figure, are lists of clients of the package, and also the list of packages upon which this package depends. There are also links to the documentation of the package, if this exists, and to a CMT tool which shows the graphic dependencies graphically.

## 4.3.  Conclusions and Future Plans

Tag Collector has become an essential tool for Atlas developers. It is the central point for obtaining information about packages. For most information, it is more convenient to browse than the CVS repository, or the afs file system.

The possibility of enforcing release policy has increased the coherence and the quality of our releases. The facility of running nightly builds augments the efficiency of development.

Centralized pointers to package documentation give us a starting point for one of our major future tasks – which is to improve the quality of our documentation in general.

It has become evident recently that the Tag Collector is almost a victim of its own success. It was developed very rapidly using PHP, and a specification produced by a limited group of people.

Regularly new features are requested, some of which are in contradiction with the original specification. In consequence, a complete new version of Tag Collector is planned for the second part of this year. This version will exploit more fully the AMI API described above, and this will allow the tool to become more flexible.